\documentclass[11pt]{article}

\usepackage{myArticle,subcaption}
\graphicspath{{../figures/}}
\usetikzlibrary{decorations.pathreplacing,arrows, arrows.meta}

\title{Change Point Detection in  Nonstationary Sub-Hourly Wind Time Series}
\author[1]{Sakitha~Ariyarathne\thanks{sariyarathne@smu.edu}}
\author[1]{Harsha~Gangammanavar\thanks{harsha@smu.edu}}
\author[1]{Raanju~R.~Sundararajan\thanks{rsundararajan@mail.smu.edu}}
\date{First submission: May, 2021.}

\affil[1]{Department of Engineering Management, Information, and Systems, Southern Methodist University, Dallas TX}

\newif\ifpaper
\papertrue

\begin{document}\thispagestyle{empty}
\maketitle

\begin{abstract}
In this paper, we present a nonparametric method for detecting change points in multivariate nonstationary wind speed time series. The change point detection method identifies changes in covariance structure and decomposes the nonstationary multivariate time series into stationary segments. We also present parametric and nonparametric approaches to model and simulate new time series within each stationary segment. The proposed simulation approach retains the statistical properties of the original time series and therefore, can be employed for simulation-based analysis of power systems planning and operations problems. We demonstrate the capabilities of the change point detection method through computational experiments conducted on wind speed time series at five-minute resolution. We also conduct experiments on the economic dispatch problem to illustrate the impact of nonstationarity in wind generation on conventional generation and location marginal prices. 
\end{abstract}

\section{Introduction} \label{sect:intro}
There is a growing need for a sustainable solution to problems attributed to pollution and climate change. Electricity production is one of the most concerning sectors as the largest contributor to greenhouse gas emissions from the use of fossil fuels. Moreover, the amount of fossil fuels is depleting resulting in higher costs \cite{PEREAMORENO2017624} and adding to the energy security concern for nations around the world.  Renewable energy resources such as geothermal, hydroelectric, wave, tidal, biofuel, solar thermal, and wind have been gaining a lot of attention since the late twentieth century as viable and environmentally friendly solutions to this energy crisis. 

Wind energy is one of the most economical renewable energy resources \cite{kumar2016wind}. In the United States, electricity production from wind resources contributes to over $7.3\%$ of the total electricity generation. However, the integration of a large amount of wind energy resources in power systems poses several operational and reliability challenges. These challenges are attributed to the inherently stochastic nature of wind generation. In addition to uncertainty, the wind energy resources also exhibit large fluctuation in their output in a short duration of time. Such stochastic and intermittent nature in the generation is also observed in solar generation. These characteristics also exacerbate the difficulty to obtain accurate forecast of wind and solar generation. To address this issue, system operators are forced to incorporate large amounts of operating reserves in the power network \cite{NREL2011}. The use of expensive operating reserves results in an overall increase in operational costs negating the cost benefits of renewable wind and solar generation. 

To manage renewable energy resources, power systems planning and operations problems are often modeled using the principles of stochastic programming (SP). This modeling approach results in optimization models that are stated with expectation-valued objective function and/or stochastic constraints. Given the large-scale nature of these problems, the SP-models are solved using approximation methods that utilize scenarios of uncertain parameters such as renewable generation and electricity demand in lieu of exact distributional information. The use of intermittent resources has also pushed the system operators to use fine timescale models of planning and operations problems with resolutions as low as $5$ minutes \cite{CAISO2013}. The use of SP models at such fine resolution has the potential of improving system operations as shown in \cite{Gangammanavar2016stochastic}. However, these benefits can be realized only if there are models of renewable generation that are capable of efficiently generating scenarios while retaining the statistical properties of the underlying stochastic processes.

The determining element of wind energy generation is wind speed. Since this is a natural phenomenon, factors that affect wind speed are not easy to recognize. Therefore, modeling wind speed accurately has been a well-researched area among both statistical and power systems communities \cite{wu2007literature}. Wind speed modeling methods in the literature can be classified into five categories. 1) Persistence methods; these models assume that the present wind speed stays consistent in the future. That is, wind speed at time $t$ can be used as the wind speed at $t + \Delta t$. Even though this technique outperforms most of the other sophisticated models for very short-term predictions, its performance diminishes very quickly as the prediction length increases. 2) Physics-based methods; these models utilize Numerical Weather Prediction (NWP) data like temperature, pressure, surface roughness, precipitation, and geographical obstacles to predict short-term wind speed. In stable weather conditions, these models can produce accurate predictions. However, the high computational burden that comes with solving these complex mathematical models limits their usefulness in real-time power systems operations. 3) Artificial Intelligence (AI) methods; these models use a large amount of historical data to identify patterns between input features and wind speeds. These techniques include Artificial Neural Networks (ANN) \cite{khodayar2018spatioNN, shukur2015dailyKF}, Support Vector Machine (SVM) \cite{wang2015robust}, multi-layered perceptron \cite{campbell2005novelNN}, among others. The prediction power of these models also diminishes quickly with the prediction time horizon. 4) Statistical methods; these are data-driven models
that are easily applicable, interpretable and yield computationally efficient  methods that can produce accurate short-term predictions. The statistical methods include time series analysis techniques such as Autoregressive (AR), Autoregressive Moving Average (ARMA), and autoregressive integrated moving average (ARIMA) models \cite{brockwell_davis}. An ARMA model has been used in \cite{wangdee2006consideringARMA} to simulate wind speed scenarios using Monte Carlo simulation. A time-shifting technique has been used in \cite{xie2009consideringTimeShifting} to simulate wind speeds using an AR model. Hourly wind speed data was modeled using an ARMA model in \cite{torres2005forecastNavarre}. A vector AR model (VAR)  \cite{HelmutVAR} has been developed using spatio-temporal data of several nearby wind farms in \cite{de2005predictiveSparse}. 5) Hybrid methods; Any combination of the above approaches can be classified as a hybrid method. 

In this study, we focus on applying statistical methods to model wind speed time series. Our choice is motivated by the fact that our statistical models are supported by sound theoretical properties while retaining computational requirements within acceptable limits. However, there are few obstacles to applying statistical methods to wind time series. Wind farms located in close proximity share common spatio-temporal properties. The statistical models employed for power systems problems have predominantly involved univariate analysis. Such time series models (e.g., AR, ARMA, and ARIMA) fail to capture the spatial relationships. Therefore, the first challenge pertains to capturing spatio-temporal information by treating wind speed time series from nearby locations as a single multivariate time series. Such models have been employed in power systems, for instance, \cite{miranda2007spatially, zhao2018correlation} present vector autoregressive (VAR) models. 

When either univariate or multivariate time series analysis is employed, the standard modeling approaches, including those mentioned above, largely require stationarity of the underlying stochastic process. However, for wind speed data, there is no guarantee that this assumption is satisfied. The most common approach to address the requirement of stationarity is to standardize the original data using the mean and the variance of the entire time series \cite{papavasiliou2015stochasticFiltering, brown1984timeStand}. This approach  does not make the time series stationary in terms of variance and temporal dependence. This leads to statistical inconsistencies between the original time series and those simulated based on models built using standardized data.  Another way of addressing nonstationarity is to segment the entire series into stationary pieces. However, this segmenting cannot be done arbitrarily. The traditional way of identifying points where an abnormality occurs is using control charts. However, the approach is more suitable in finding isolated major changes in an ongoing process.

Finally, all the parametric time series methods require certain distributional assumptions. That is, to apply these methods, the data, or at least the residual or error terms obtained after removing the trend and seasonal components, is required to follow a Gaussian distribution. For instance, \cite{friedrich1997descriptionPhy, ragwitz2001indispensablePhy} claim that wind speed data is more likely to follow an approximately Gaussian distribution as the time resolution of the time series increases. Further, Weibull distribution can be used to estimate parameters of an approximate-Gaussian distributed wind speed data \cite{hennessey1977someWeibull, morales2010methodologyWeibull}. To address the distributional requirement, previous works (e.g., \cite{papavasiliou2015stochasticFiltering})  propose to filter the data such that the final sample has a Gaussian distribution.

In light of these observations, 
\ifpaper
the main contributions of this paper are as follows.
\fi
\begin{enumerate}
    \item \emph{Change point detection.} We propose a nonparametric change point detection method for wind speed data. Our approach is based on a recent work \cite{sundararajan2018nonparametric}, and treats the wind speed data as multivariate nonstationary time series. This data-driven technique identifies stationary segments of the entire series using change points. The change points are detected based on differences in the covariance structure measured using spectral density matrices \cite{brillinger81}. The method is nonparametric and does not make distributional assumptions. 
    \item \emph{Simulation of wind speed, parametric vs nonparametric approaches.} We discuss parametric and nonparametric time series methods for simulating sub-hourly wind speed data. These simulation methods are applied to all the stationary segments identified using the change point detection method. The proposed simulation methods retain statistical properties of the original time series and this is witnessed in the computational experiments involving the original and simulated wind speed time series. 
    \item \emph{Application to a Power Systems Operations Problem.} We demonstrate the application of time series simulated using our proposed approaches to the economic dispatch problem used for power systems operations planning. Our computational experiments illustrate that the distributional changes, at identified time points (change points), in the input wind speed time series, results in distributional changes at the same time points in time series corresponding to conventional generation and location marginal prices. Our approach provides systematic tools to analyze power systems both from the operations and market stability perspectives.
\end{enumerate}

The rest of the paper is organized as follows. In section \S\ref{sect:statModel}, we present the change point method to detect  changes within wind speed time series. We also discuss parametric and nonparametric approaches to simulate new wind speed scenarios. In section \S\ref{sec:marketClearingApp}, we introduce an  economic dispatch problem to illustrate the application of our wind speed modeling method to an important problem in power systems operations. In section \S\ref{sec:compExp}, we present the computational experiments conducted on wind time series data and the behavior of the economic dispatch problem outputs. This is followed by the conclusion in section \S\ref{sec:statConclussion}.

\section{Statistical Methodology} \label{sect:statModel}
In this section, we present the statistical methods for modeling wind speed time series. We provide a description of the change point method and then discuss the parametric and nonparametric simulation techniques. We start with the model assumption and the motivation to study wind speed data as a multivariate nonstationary time series.

Let $\{\vecB{w}_t\}_{t=1}^T$ denote the $L$-dimensional wind speed stochastic process. This process can be decomposed into three $L$ dimensional terms: trend $(\vecB{x}_t)$, seasonal $(\vecB{s}_t)$, and residual $(\vecB{r}_t)$ terms. That is, 
\begin{align} \label{eq:main_model}
    \vecB{w}_t = \vecB{x}_t + \vecB{s}_t + \vecB{r}_t, \qquad  t =1,2,\hdots,T.
\end{align}
We postpone the discussion on estimating the trend and the seasonal terms to the end of this section. Observe that model \eqref{eq:main_model} assumes presence of trend and seasonal terms for each component (or variate) of the multivariate series. The serial and contemporaneous dependence among the $L$ components of $\vecB{w}_t$ is assumed to be driven exclusively by the $L$-dimensional residual series $\vecB{r}_t$. If this series $\{\vecB{r}_t\}$ is stationary, application of conventional time series models, such as VARMA, is appropriate. The model components would  then be estimated by minimizing the least squares error \cite{HelmutVAR}. 

As an illustration, we present the correlation matrix of the residual series $\vecB{r}_t$ in Table  \ref{tab:corMatResFeb19} for a randomly selected day (Feb. $19$, 2011\footnote{Data source: https://www.nrel.gov/grid/wwsis.html}). The $L=5$ component series here are sub-hourly wind speed data gathered at 5 wind farm locations in Oklahoma. Note that the presence of trend and/or seasonal characteristics in the original series $\vecB{w}_t$ makes it inappropriate for an analysis using the standard correlation coefficient and hence we only present the correlation results based on the residual series $\vecB{r}_t$. Furthermore, the model assumption in \eqref{eq:main_model} implies that the dependence among components of $\vecB{w}_t$ is captured by dependence among components in $\vecB{r}_t$. The results from Table \ref{tab:corMatResFeb19} indicate a strong linear relationship among individual components of the multivariate series $\vecB{r}_t$, suggesting a multivariate consideration of the time series.  

\begin{table}[h!]
\centering
\renewcommand{\arraystretch}{1.5}

 \caption{Correlation matrix of the residual series  $\vecB{r}_t$ (Feb. 19, 2011).}{
    \begin{tabular}{lrrrrr}
    \hline
          & \multicolumn{1}{l}{Series 1} & \multicolumn{1}{l}{Series 2} & \multicolumn{1}{l}{Series 3} & \multicolumn{1}{l}{Series 4} & \multicolumn{1}{l}{Series 5} \\
    \hline
    Series 1 & 1.00 & 0.55 & 0.60 & 0.72 & 0.52 \\
    Series 2 & 0.55 & 1.00 & 0.43 & 0.29 & 0.14 \\
    Series 3 & 0.60 & 0.43 & 1.00 & 0.67 & 0.62 \\
    Series 4 & 0.72 & 0.29 & 0.67 & 1.00 & 0.55 \\
    Series 5 & 0.52 & 0.14 & 0.62 & 0.55 & 1.00 \\
    \hline
    \end{tabular}
    \label{tab:corMatResFeb19}}
\label{tab:corMatFeb19}
\end{table}

Next, an application of a multivariate stationarity test \cite{subbarao15} on the residual series $\vecB{r}_t$ suggests that the stationary assumption does not hold for every wind speed time series. As an illustration, we present the p-values from this test on hourly and sub-hourly wind speed data gathered at the same $L=5$ wind farm locations in Oklahoma. The analysis here was conducted on day-long sub-hourly wind speed data for five randomly selected days and week-long hourly wind speed data\footnote{The choice of selecting $7$ consecutive days for hourly data is to ensure that sufficient data is available for model parameter estimation. The hourly data has $24 \times 7 = 168$ data points and sub-hourly data has $12 \times 24 = 288$.} starting on the same randomly selected days. The test checks for covariance stationarity of the multivariate time series with a null hypothesis that the given multivariate series is stationary. The p-values in the table provide a strong evidence to reject the null hypothesis, leading us to conclude that both sets of sub-hourly and hourly time series exhibit nonstationarity. 
\begin{table}[h!]
    \centering
    \caption{The p-values from multivariate stationarity test for the five wind farm locations in Oklahoma.}
    \renewcommand{\arraystretch}{1.5}
    \begin{tabular}{ccc}
        \hline
         Day &  Hourly & Sub-hourly\\
         \hline
         Feb. 19, 2011 & $1.211e-03$ & $<1e-06$\\
         Apr. 21, 2011 & $8.633e-04$ & $<1e-06$ \\
         Aug. 07, 2011 & $3.185e-04$ & $<1e-06$\\
         Nov. 01, 2011 & $8.069e-06$ & $<1e-06$\\
         Dec. 04, 2011 & $1.347e-06$ & $<1e-06$\\
         \hline
    \end{tabular}
    \label{tab:stationaryTestPval}
\end{table}

\begin{figure*}
    \centering
    \begin{subfigure}[t]{\textwidth}
    \includegraphics[width=0.99\textwidth]{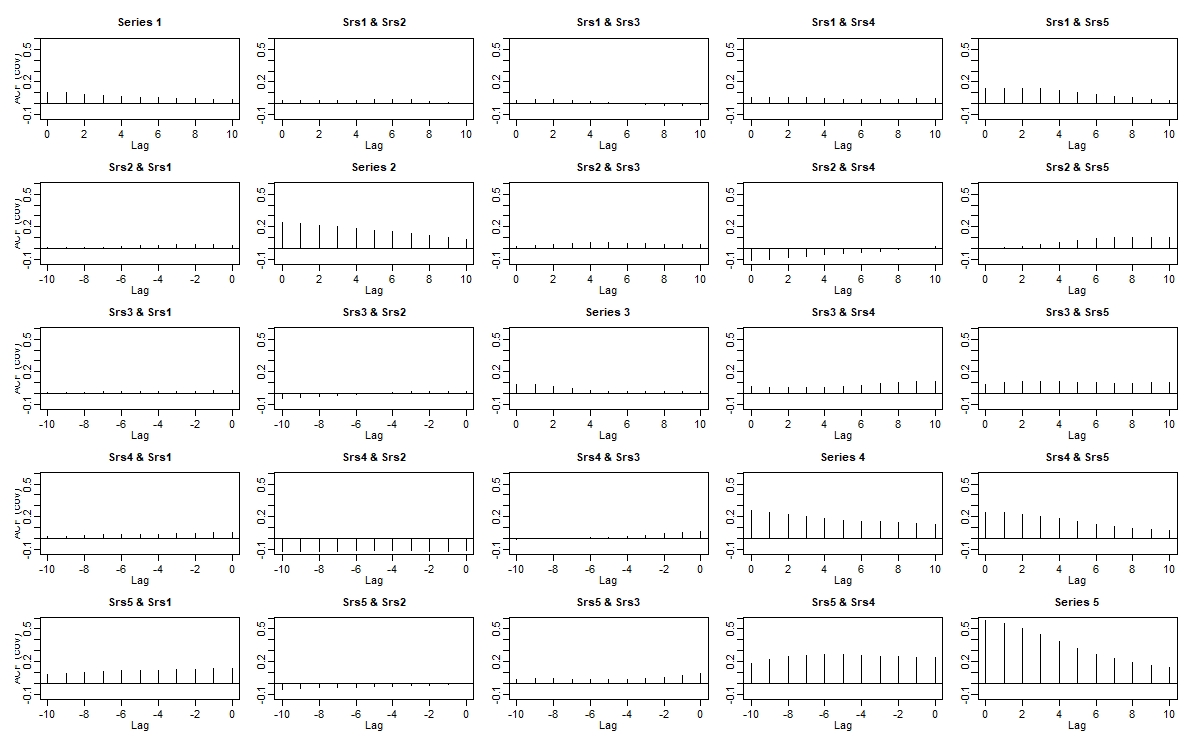}
    \caption{First 100 observations.}
    \label{fig:ACFsubHourlyfirst100feb19}
    \end{subfigure}
    \begin{subfigure}[t]{\textwidth}
    \includegraphics[width=0.99\textwidth]{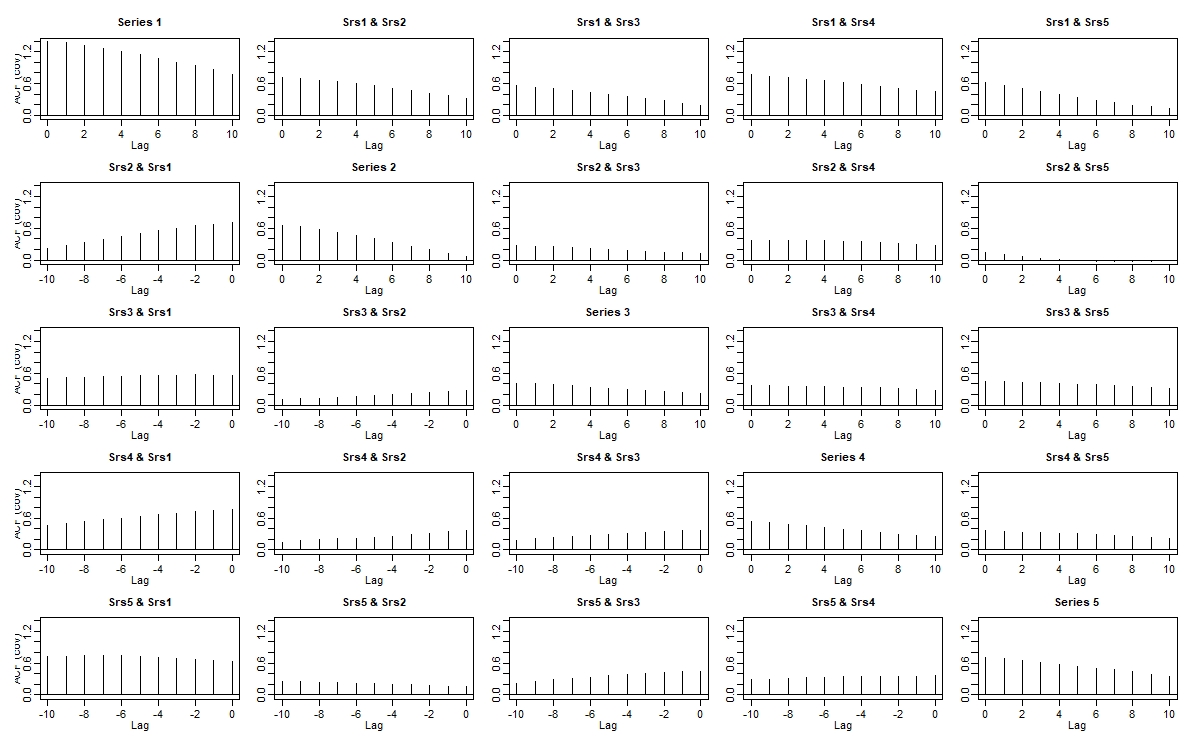}
    \caption{Last 100 observations.}
    \label{fig:ACFsubHourlylast100feb19}
    \end{subfigure}
    \caption{Autocovariance plots of sub-hourly residual data for February $19^{th}$.}
    \label{fig:ACFsubHourlyFeb19}
\end{figure*}

Nonstationarity in the wind speed time series can also be witnessed by checking the autocovariance over different segments in time. Figure \ref{fig:ACFsubHourlyFeb19} depicts the changes in the covariance structure of sub-hourly data recorded on February $19$ over time. In Figure \ref{fig:ACFsubHourlyfirst100feb19}, the covariance patterns of the first 100 observations are entirely different form the last 100 observations in Figure \ref{fig:ACFsubHourlylast100feb19}. For instance, the first 100 observations of series 1 and 3 (Figure \ref{fig:ACFsubHourlyfirst100feb19}) suggest a small relation over different lags while the last 100 observations (Figure \ref{fig:ACFsubHourlylast100feb19}) suggest otherwise. Even the univariate plots along the diagonal in both figures \ref{fig:ACFsubHourlyfirst100feb19} and \ref{fig:ACFsubHourlylast100feb19} show their variance structure change over time. Similar changes were also observed in the covariance structure for sub-hourly wind speed data for other days listed in Table \ref{tab:stationaryTestPval} as well as the hourly wind speed data. This highlights the prominence of covariance nonstationarity in wind speed data. These preliminary analyses not only emphasize the need to pursue multivariate analysis, but also illustrate that the direct application of stationary VARMA models is not suitable due to inherent nonstationarity. 

\subsection{Change Point Detection} \label{sec:change_point}
Change point analysis identify changes in the statistical properties of nonstationary time series data. Applications of change point analysis are widespread; see \cite{changePointApp1, changePointApp2, changePointApp3, changePointApp4} for examples. Change point analysis can be performed in parametric \cite{parametricChangePoint} and nonparametric \cite{changePoints} settings. Parametric methods assume the data arises from a certain distribution and change points are obtained by tracking changes in the parameters of that distribution across time. Nonparametric methods work without such assumptions and we resort to this approach to identify change points in wind speed data.

 A time point $\tau \in (1,T)$ is regarded as a change point when the statistical properties of the series $\vecB{r}_t$ on either side of that time point $\tau$ are different. For instance, if the covariance $Cov( \vecB{r}_t , \vecB{r}_{t+h} )$ for $t \in (\tau -\delta , \tau]$ at some lag $h \geq 0$ is  different from $Cov( \vecB{r}_t , \vecB{r}_{t+h} )$ for $t \in (\tau,\tau+\delta]$ for some neighborhood length $\delta>0$, then $\tau$ is a change point. The same concept can be extended to $M > 1$ change points at $\tau_1, \tau_2, \cdots , \tau_M \in (1,T)$ and we  let $\tau_0 = 0$ and $\tau_{M+1} = T$. In this case, the original stochastic process can be split into $M+1$ \textit{ stationary segments} indexed by $I^k: = (\tau_k, \tau_{k+1}]$, for $k=0,1,\hdots,M$. Notice that the above procedure requires the knowledge of the number and locations of change points $\tau_k$, for $k = 1,\ldots,M$. In some prior works, the number of change points is assumed to be known and their locations are estimated \cite{polansky2007detectingKnownNumber}.

In many practical settings, however, it is unreasonable to expect that either the number or the location of change points is known a priori. We resort to a nonparametric method for detecting changes in the covariance structure of the residual time series $\vecB{r}_t$. Our approach is based on the recently developed nonparametric method in \cite{sundararajan2018nonparametric}. We provide a brief overview of this approach here and refer the reader to \cite{preuss2015detection, matteson2014nonparametric, sundararajan2018nonparametric} for detailed exposition and analysis. 

Let $f( s , \omega)$ for $\omega \in \lbrack -\pi, \pi \rbrack$, be the $L \times L$ spectral density matrix of the residual series $\vecB{r}_t$ at frequency $\omega$ and time $s$. Note that the spectral matrix, at different frequencies, carries all the information about the covariance structure of the series $\vecB{r}_t$. The key tool for locating change points is a deviation metric that serves as a measure of departure from stationarity. More precisely, for any candidate change point $\tau$, we consider $D(\tau)$ as the integrated squared Euclidean norm of the vectorized difference of the spectral density matrices in a neighborhood around point $\tau$. We have
\begin{align} 
    D(\tau) = \cfrac{1}{2\pi} \int_{-\pi}^{\pi} || \text{vec} \Big( f_L(\tau,\omega) - f_R(\tau, \omega) \Big) ||^2 d\omega. \label{eq:distanceMetric}
\end{align}
where, $f_L(\tau, \omega)$ and $f_R(\tau,\omega)$ are the spectral matrices evaluated over the neighborhoods ($\tau-\delta, \tau$) and ($\tau, \tau + \delta$), respectively. Here, $\delta>0$ is the size of the local neighborhood and $\omega \in \lbrack -\pi , \pi \rbrack$ is the frequency. While estimating $D(\tau)$, we consider $N$ observations on either side of the candidate change point $\tau$ and this $N$ serves as a proxy for $\delta$. Then, an estimate of $f_L(\tau,\omega)$ can be defined using the discrete Fourier transform (DFT) and the periodogram of the series $\vecB{r}_t$:
\begin{align}
 J_L(\omega)= \frac{1}{\sqrt{2\pi N}} \sum_{s = \tau - N + 1}^{\tau} \vecB{r}_s e^{-it\omega},\quad I_L(\omega)=J_L(\omega)J_L(\omega)^*,
\end{align}
where $J_L(\omega)^*$ is the conjugate transpose of $J_L(\omega)$. The estimated $L \times L$ spectral density matrix for $\omega \in [-\pi, \pi]$ is given by
\begin{align}
    \hat{f}_L(\omega) = \cfrac{1}{N} \sum_{j=-\lfloor \frac{N-1}{2} \rfloor}^{\lfloor \frac{N}{2} \rfloor} K_h(\omega - \omega_j)I_L(\omega_j),
\end{align}
where $\omega_j=\frac{2\pi}{N}j$ and $K_h(\cdot)=\frac{1}{h}K(\frac{\cdot}{h})$. Here, $K(\cdot)$ is a non-negative symmetric kernal function and $h$ is the bandwidth. An estimate for $f_R(\tau , \omega)$ can be obtained similarly. Finally, an estimate for $D(\tau)$ in \eqref{eq:distanceMetric} can be defined as
\begin{align} \label{eq:d_hat}
    \hat{D}(\tau) = \cfrac{1}{2\pi} \int_{-\pi}^{\pi} || \text{vec} \Big( \hat{f}_L(\tau,\omega) - \hat{f}_R(\tau, \omega) \Big) ||^2 d\omega.
\end{align}
The method operates in a sequential manner wherein the point with maximal value in $D(\cdot)$ is identified first and tested for significance. To test the significance of this candidate change point at level $\alpha$, the estimate in \eqref{eq:d_hat} serves as a test statistic. The method then moves to the time point with the next biggest value in $D(\cdot)$. The significant change points are identified sequentially until there are no more time points deemed as change points by the test.  This procedure yields the number $M$ and location of the  change points. 

Once we identify the $M$ change points, we are left with $M+1$ stationary segments and the time series within these segments can be treated as stationary series. Our goal then is to mimic this segmentation by simulating stationary time series in each of these segments, thereby producing a realization of the entire series that is statistically consistent with the original time series. Producing accurate simulations of the original series helps greatly in power systems operations and this application is discussed in Section \ref{sec:marketClearingApp}. Next in Sections \ref{sec:parametric_sim} and \ref{sec:nonp_sim}, we present two methods to simulate from a stationary time series. These are the methods that will be used to simulate time series data within each of the $M+1$ identified segments. 

\subsection{Simulating stationary processes: Parametric method}\label{sec:parametric_sim}
Modeling a stationary time series can be achieved using the  \textit{Vector Autoregression} (VAR) model. The $p^{th}$ order VAR model for the residual vector $\vecB{r}_t$ in segment $I^k: = (\tau_k, \tau_{k+1}]$, for $k=0,1,\hdots,M$, can be written as
\begin{align}
    \vecB{r}_t = W_{1,k} \vecB{r}_{t-1} + \dots + W_{p,k} \vecB{r}_{t-p} + \Psi_{t,k} \label{generalVAR},
\end{align}

where $W_{1,k} , W_{2,k} , \cdots, W_{p,k} $ are the $L \times L$ coefficient matrices, and $\Psi_{t,k}$ is the $L$-dimensional noise term. Selecting the model order $p$ is an important step in VAR model estimation and can be achieved through selection methods such as Akaike information criterion (AIC) or Bayesian information criterion (BIC) \cite{HelmutVAR}. 

After fitting VAR models in each of the $M+1$ stationary segments, the fitted models can be used to simulate stationary series for each segment, thereby resulting in a realization for the entire series \cite{kreiss1992}.

The parametric bootstrap method described above requires specification of the distribution of the noise term $\Psi_{t,k}$. In finite sample situations, fitting VAR models, especially those with high model orders $p$, can be a challenging estimation problem. In the next section, we describe a nonparametric approach to simulating from a stationary time series. This approach avoids fitting parametric models such as VAR and its associated problems such as model order estimation and error term distribution specification. 

\subsection{Simulating stationary processes: Nonparametric method} \label{sec:nonp_sim}

The block bootstrap technique is one of the nonparametric methods that can be used to simulate from a stationary time series \cite{politis94}. In this method, blocks of the original time series are created using a block length parameter. The simulated series is obtained by resampling blocks of observations as opposed to resampling individual observations. The block length is an important parameter choice and the technique in \cite{politis04} can be utilized. Though the previous method is developed for univariate time series, one can obtain optimal block lengths for each component series in the multivariate series and compute an average block length at the end. The block bootstrap method does not require the distribution of the error terms, avoids small sample estimation problems in parametric models such as VAR, and requires selecting the one block parameter. In practice, the choice between a parametric or nonparametric simulation technique depends on the length of the $M+1$ segments discovered by the chagne point method in Section \ref{sec:change_point} and also the estimated model order $p$ in the VAR bootstrap described in Section \ref{sec:parametric_sim}. Parametric resampling methods face estimation issues when there are many parameters to be estimated with few observations and this is further discussed in Section \ref{sec:performance_energy}. 

\subsection{Estimating the trend and seasonal terms} \label{subsec:trendSeasonalEstimate}

Our model assumption in \eqref{eq:main_model} includes a trend term $\vecB{x}_t$ and a seasonal term $\vecB{s}_t$. Popular approaches to estimating the trend term $\vecB{x}_t$ are smoothing techniques such as moving average, exponential smoothing, and local least squares. The sub-hourly and hourly wind speed data do not exhibit linear nor monotone trend behavior. We thus utilize the \textit{Loess regression algorithm} \cite{LOESS} to estimate the trend term wherein the algorithm fits regression models over local time windows.

The seasonal component $\vecB{s}_t$ is the repeating pattern of a time series at regular intervals. After removing the trend term, plots of the spectral density \cite{brillinger81} and autocorrelation  of the univariate components of $\vecB{s}_t$ provide an initial idea of presence of seasonality in the time series. To identify seasonal periods and estimate the seasonal term $\vecB{s}_t$, one can resort to the additive models suggested in \cite{tbats}. In modeling sub-hourly wind speed data over individual days, seasonality is not an issue due to the small time period under consideration.  

The entire procedure of modeling and simulating wind speed time series is summarized in Algorithm \ref{alg:procedure}. The procedure accepts a $L$-dimensional time series as input, prepossesses the time series (Step 1), identifies the change points at user defined significance level $\alpha$ (Step 2) and returns the desired number of simulated time series (Step 4). The simulated series are generated using either parametric or nonparametric approaches (Step 3).
\begin{algorithm}[h!]
\caption{Simulating wind speeds} \label{alg:procedure}
\begin{algorithmic}[1]
\State {\bf Input:}  $L$ dimensional time series from nearby wind farm $\vecB{w}_t, ~t=1,2,\hdots,T$; significance level $\alpha$; number of simulations $N$.
\State {\bf Step 1:} Identify trend $\vecB{x}_t$ and seasonal $\vecB{s}_t$  components. Obtain residual series $\vecB{r}_t = \vecB{w}_t - \vecB{x}_t - \vecB{s}_t$. (\S\ref{subsec:trendSeasonalEstimate})
\State {\bf Step 2:} Run the change point detection method on $\vecB{r}_t$ for the input significance level $\alpha$ to obtain change points $\tau_0, \tau_1, \ldots, \tau_{M+1}$ where $\tau_0 = 0$ and $\tau_{M+1} = T$. Separate $\vecB{r}_t$ into segments indexed by $I^k:=(\tau_k,\tau_{k+1}]$ (\S\ref{sec:change_point})
\State {\bf Step 3:} Obtain time series model for each segment.
	\For{$k = 0,\ldots,M$}
		\State Use parametric method, \textit{VAR(p)}  to obtain model for the segment $I^k$. (\S\ref{sec:parametric_sim})
		\If{$p<5$}
		    \State Simulate $N$ time series using the \textit{VAR(p)} model.
		\Else
			\State Use nonparametric method, block bootstrap to obtain $N$ simulated time series. (\S\ref{sec:nonp_sim})
		\EndIf
	\EndFor
\State {\bf Step 4:} Append all the simulated segments to obtain a time series of length $T$. Add trend $\vecB{x}_t$ and seasonal $\vecB{s}_t$ components. This is done separately for each of the $N$ simulations. 
\State {\bf Output:}  $N$ simulated wind speed time series.
\end{algorithmic}
\end{algorithm}

\section{Application to Economic Dispatch Problem} \label{sec:marketClearingApp}
With large-scale integration of renewable energy resources for electricity generation, the planning and operations problems in power systems face several challenges that stem from the inherent stochasticity in these resources. The deterministic optimization tools employed by the system operators raise several concerns regarding reliable operations as well as consistency in electricity prices. Stochastic programming 
provides a suitable framework to address both these issues when optimization model parameters are uncertain. However, the effect of nonstationarity on system operations has not be investigated before. In light of this, we present a critical power systems operations problem to illustrate how nonstationarity in input parameters (wind speed time series) impacts optimization output (conventional generation and electricity prices).

Consider a power system with the set of lines $\set{L}$, buses $\set{B}$, generators $\set{G}$, and load $\set{D}$. The following mathematical program is a deterministic quadratic programming model build to solve the economic dispatch problem. The decision variables of the model are as follows: $g_i$ is the generation of $i \in \set{G}$ generator, $d_i$ is the demand met at $i \in \set{D}$ load, $f_{ij}$ is the power flow on line $(i,j) \in \set{L}$, and $\theta_i$ is the angle at bus $i \in \set{B}$. The model is given by
\begin{subequations}
\begin{align}
  	\min~& \sum_{i \in \mathcal{G}} a_ig_{i}^2 + b_ig_i   -  \sum_{i \in \mathcal{D}} \beta_{i} d_{i} \label{eq:statobj}\\
  	\text{s.t.}~ &\sum_{j:(j,i) \in \mathcal{L}} f_{ji}  - \sum_{j:(i,j) \in \mathcal{L}} f_{ij}  + \sum_{j \in \mathcal{G}_i} g_{j} - \sum_{j \in \mathcal{D}_i} d_{j}  = 0 \notag \\
  	&\hspace{5cm} \forall i \in \mathcal{B} ,  \label{eq:statflowBalance} \\
  	& \underline{\Delta}_i + g_i^\star \leq g_i \leq \overline{\Delta}_i + g_i^\star ~~~ \forall i \in \mathcal{G}, \label{eq:statramping} \\
  	& f_{ij}  = \cfrac{V_iV_j}{X_{ij}}(\theta_{i} -\theta_{j} ) \quad \forall (i,j) \in \mathcal{L}, \label{eq:statkirchhoffTherom} \\
    &  \underline{G}_{i} \leq  g_{i} \leq \overline{G}_{i} \quad  \forall i \in \mathcal{G}, \label{eq:statgenBounds} \\
    & 0 \leq d_{i}  \leq {D}_{i} \quad \forall i \in \mathcal{D}, \label{eq:statdemBounds}\\  
    &  \underline{F}_{ij} \leq  f_{ij} \leq \overline{F}_{i} \quad  \forall (i,j) \in \mathcal{L}. \label{eq:statlineBounds}
\end{align}
\label{eq:statModelClearingModel}
\end{subequations}
In the above, $a_i$ and $b_i$ are the cost coefficients, $\underline{G_i}/\overline{G_i}$ are the production capacity limits, and $\underline{\Delta_i}/\overline{\Delta_i}$ are the ramp-down/ramp-up limits of the generator $i \in \set{G}$. The bidding price and demand of the load $i \in \set{L}$ is denoted by $\beta_i$ and $D_i$, respectively. $V_i$ is the voltages at bus $ i \in \set{B}$; $X_{ij}$ is the line impedance and $\underline{F}_{ij}$/$\overline{F}_{ij}$ are the line capacities of the line $(i, j) \in \set{L}$.

The objective of the model, \eqref{eq:statobj} minimizes the total operating cost of the entire power system. The first term represents the total power generating cost and the second term represents the total income from the consumers. The first constraint, \eqref{eq:statflowBalance} captures the power flow balancing constraints. That is, the total power inflow of a bus $b \in \set{B}$ is equal to the total outflow. The second set of constraints,\eqref{eq:statramping} are the ramping limitations of generators in the system and the third set, \eqref{eq:statkirchhoffTherom} represent the direct-current (linear) approximation of power flow on line $(i,j) \in \set{L}$. Constraints \eqref{eq:statgenBounds}, \eqref{eq:statdemBounds}, and \eqref{eq:statlineBounds} are the bounds on the respective decision variables. The optimization model is stated for generation amount $(g_i^\star)_{i \in \set{G}}$ for the previous time period and affects the current generation amount through the ramping constraints \eqref{eq:statramping}. Notice that, \eqref{eq:statModelClearingModel} is a quadratic program with affine equality constraints and bounded variables. 

Since the input to the optimization model follows a stochastic process, its output, the optimal primal and dual solutions, can also be viewed as stochastic processes. To capture these output stochastic processes, the model is solved iteratively in a rolling-horizon manner. That is, at time period $t$, a model instance with wind generation corresponding to wind speed $\mathbf{w}_t$ is setup and solved. The resulting optimal generation amount (denoted $g_{it}^\star$ for $i \in \set{G})$, along with wind generation amount corresponding to $\mathbf{w}_{t+1}$ is used to setup the next instance the model at time period $t+1$. The process is repeated until the end of the horizon. At a given time period $t$, the optimal dual solution of \eqref{eq:statflowBalance}, denoted by $\pi_{it}^\star$, is the location marginal price at bus $i \in \set{B}$. In our computational study, we investigate the behavior of generation $\{\mathbf{g}_t\}$ and price $\{\boldsymbol{\pi}_t\}$ stochastic processes.

\section{Computational Experiments} \label{sec:compExp}
In this section, we present the numerical results of of the change point detection method presented in \S\ref{sec:change_point}. The experiments were conducted on five wind farm locations in Oklahoma. The data was collected from the National Renewable Energy Laboratory (NREL) website (https://www.nrel.gov/). All the experiments were conducted on Intel Core i3 CPU 2.2 GHz processor and 8 GB RAM with SSD storage. All the statistical modeling and the simulations were implemented using R, version 3.5.1, the optimization model was implemented in C++, and solved using CPLEX 12.9.

The five wind farm locations considered are in close proximity to maintain the spatio-temporal properties, as demonstrated in \S\ref{sect:statModel}.  Five minutes apart sub-hourly data was gathered for each location. Figure \ref{fig:OriginalTSSep15} depict wind speed time series data for the selected five wind farms for September $15^{th}$ 2011. The speeds are varying from $0.613$ ms$^{-1}$ to $15.014$ ms$^{-1}$. These wind speeds result in energy generation that range between $0$ MWh to $21.02$ MWh. 

\subsection{Change point detection}
To illustrate the potential of the change point detection method presented in \S\ref{sect:statModel}, we conducted experiments on different sets of multivariate wind speed time series based on different choices of length of series and number of variables. Sub-hourly data for $12$, $24$, and $48$ hour time periods were obtained using $T = 144$, $288$ and $576$ data points of wind speed  time series starting at midnight of September 15, 2011. The variates or components of the multivariate time series correspond to wind farm locations selected at random from the available dataset. We use two, three, four, and five dimensional time series in our experiments.

\begin{figure}[!h]
    \begin{subfigure}[b]{\columnwidth}  \includegraphics[width=\textwidth]{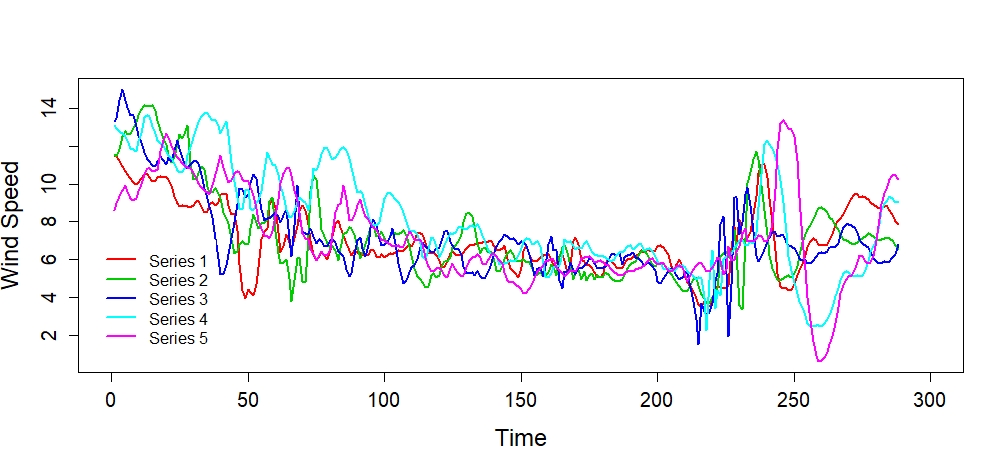}
    \caption{}  
    \label{fig:OriginalTSSep15}
    \end{subfigure}
    \begin{subfigure}[b]{\columnwidth} \includegraphics[width=\textwidth]{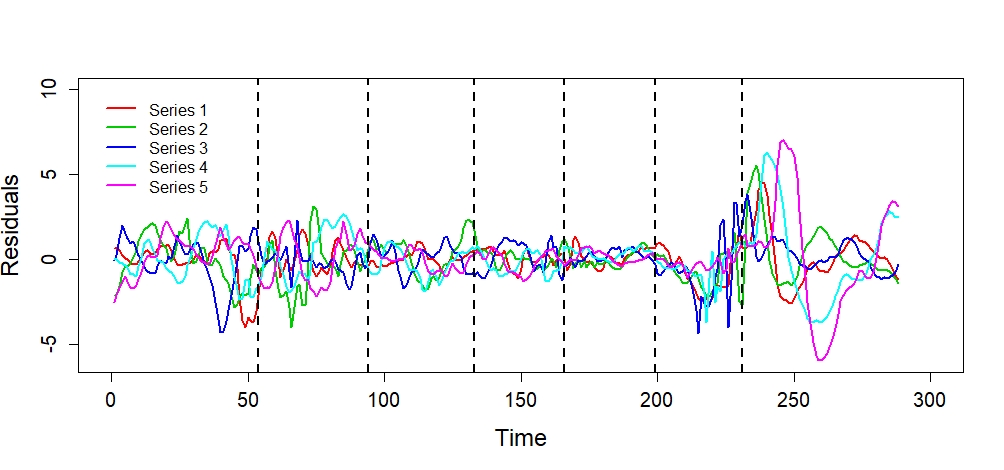}
    \caption{}
    \label{fig:ResidualTSSep15}
    \end{subfigure}
    \caption{(a). Observed wind speed series at five locations. (b). Residual series. September 15, 2011.}
    \label{fig:TSSep15}
\end{figure}

 The residuals series were obtained by removing the trend from the original series, using the Loess algorithm, for each component wind speed series. The residual time series are depicted in Figure \ref{fig:ResidualTSSep15}.  The change point method was then applied on this residual series and the results are summarized in Table \ref{tab:parameterTuning}. For instance, the change point method detected six change points at $t = 54, 94, 133, 166, 199,$ and $231$ in the 5-dimensional sub-hourly time series. These change points are identified using the vertical lines in Figure \ref{fig:ResidualTSSep15}.


A critical parameter for the change point detection method is the significance level $\alpha$. A smaller value of $\alpha$ imposes a stringent requirement for determining a candidate point to be affirmed as a change point. We recommend using $\alpha = 0.05$ or $\alpha = 0.01$ for wind speed time series. Table \ref{tab:parameterTuning} presents the results for different sets of time series at these two significance levels. The impact of the choice of significance level was revealed in the results for $12$-hours long, two-dimensional sub-hourly time series. When $\alpha = 0.05$, the change point method detected three change points at $58, 75$, and $102$. However, the chance point at $t = 102$ was not significant at $\alpha = 0.01$. Similar observation was also made for 7 days long, 5-dimensional hourly time series.


The length of the series also has an impact on the change point detection method. Table \ref{tab:parameterTuningSubHourly} shows that when series length increases the number of change points identified increases. For instance, the $12$ hours long, three-dimensional time series has two change points at $\alpha = 0.01$, whereas, the $24$ hours long series has four change points, and $48$ hours long series has five change points.  
 
\begin{table*}[!t]
\centering
\renewcommand{\arraystretch}{1.5}
\caption{Change point analysis: Series length, significance level, number of  series. Starting day: September $15^{th}$, 2011.}
\begin{subtable}{\textwidth}
\caption{Sub-hourly time series.}
  \resizebox{0.99\textwidth}{!}{
    \begin{tabular}{rlrrrr}
    \hline
       Significance &  Length of &  \multicolumn{4}{c}{Number of time series} \\
        \cline{3-6}
          level($\alpha$) & the series & \multirow{1}{*}{2} & \multirow{1}{*}{3} &         \multirow{1}{*}{4} & \multirow{1}{*}{5} \\
    \hline
    0.01  & 12hrs    & $58,75$   &  $58,103$     &   $42,58,83$    &   $41,58,83$  \\
          & 24hrs    & $89,224$   & $54,103,145,224$   & $89,231$   & $54,94,133,166,199,231$ \\
          & 48hrs    & $89,224,256,371,436$   & $89,224,256,341,435$ & $89,231,264,359,436$  & $89,231,302,368,435$\\
    0.05  & 12hrs    & $58,75,102$ & $58,103$  &  $42,58,83$     &  $41,58,83$  \\
          & 24hrs    & $89,224$   & $54,103,145,224$   & $89,231$   & $54,94,133,166,199,231$ \\
          & 48hrs    &  $89,224,256,371,436$    & $89,224,256,341,435$ & $89,231,264,359,436$  &   $89,231,264,359,435$\\
    \cline{1-6}
    \end{tabular}}
  \label{tab:parameterTuningSubHourly}
  \end{subtable}
  
  \vspace{5mm}
\begin{subtable}{\textwidth}
\centering
\caption{Hourly averaged time series.}
  \resizebox{0.6\textwidth}{!}{
    \begin{tabular}{rlrrrr}
    \hline
       Significance &  Length of &  \multicolumn{4}{c}{Number of time series} \\
        \cline{3-6}
          level($\alpha$) & the series & \multirow{1}{*}{2} & \multirow{1}{*}{3} &         \multirow{1}{*}{4} & \multirow{1}{*}{5} \\
    \hline
    0.01  & 7days    & $55,71$   &  $55,71$     &   $55,71$    &   $54,71$  \\
          & 14days    & $99,257$   & $99,257$   & $99,257$   & $99,164,257$ \\
          & 21days    & $99,257$   & $98,257$ & $98,257$  & $98,257$\\
    0.05  & 7days    & $55,71$ & $55,71$  &  $55,71$     &  $54,71,90$  \\
          & 14days    & $99,257$   & $99,257$   & $99,257$   & $99,164,257$ \\
          & 21days    &  $99,257$    & $98,257$ & $98,257$  &   $98,257$\\
    \cline{1-6}
    \end{tabular}}
  \label{tab:parameterTuningHourly}
  \end{subtable}
  \label{tab:parameterTuning}
\end{table*}

\subsection{Simulation of wind speeds}
The detected change points lead to the stationary segments and then parametric and nonparametric approaches from Sections \ref{sec:parametric_sim}, \ref{sec:nonp_sim} can be used for simulation. The parametric approach involves, at each segment, identifying the appropriate VAR model order, estimating the model parameters and simulating time series using the estimated model. The nonparametric approach, on the other hand, is based on block bootstrap carried out on each segment. 

For the sub-hourly time series in Figure \ref{fig:ResidualTSSep15}, the change point detection method indicates the presence of six change points. Thus, we decompose the time series into into seven stationary segments index by the intervals $[1,54], [55,94], [95,133], [134,166], [167,199], [200,231],$ and $[232,288]$. Segment-wise VAR model order $p$ was estimated using the Akaike information criterion (AIC) resulting in orders $p = 9, 6, 6, 5, 5, 5$, and $10$ respectively. The number of VAR parameters is given by $pL^2 + L$. Unless the series length in each segment far exceeds the number of parameters to be estimated, the quality of estimated parameters is suspect. For all the series considered in our experiments, this was found to be the case. We do not recommend using the parametric approach for wind speed when the model order is five or higher. In such cases, we recommend the nonparametric block bootstrap approach for simulation. The entire procedure to simulate wind speeds in summarized in Algorithm \ref{alg:procedure}.  

\subsection{Performance in economic dispatch problem} \label{sec:performance_energy}
To verify the impact of nonstationarity in wind speed time series on power systems operations, we implemented the rolling-horizon economic dispatch problem. For this experiment we considered a modified IEEE-30 test system \cite{UWPTest}. The system comprises of $30$ buses, $41$ lines with a maximum capacity of $50MWh$, a total of $6$ generators, and $21$ loads. We considered hourly changes in the demand at each load. 

We considered generators at bus $8$ and bus $13$ to be wind generators located on farms that share similar wind patterns. From the available sub-hourly wind speed data, two locations were randomly picked to simulate the wind speed time series. The correlation between the selected locations' original wind speed time series is $0.7265$. The change point analysis detected three change points for the bivariate time series at $t = 87, 218, 250$, resulting in four stationary segments. When parametric modeling was used, the order of the VAR model for each stationary section was $10, 8, 6,$ and $5$. Since these values are higher than the recommended model order of $5$, the nonparametric approach was used for simulation. 

\begin{figure}[!h]
\begin{subfigure}[b]{\columnwidth}
    \includegraphics[width=\textwidth, height=5cm]{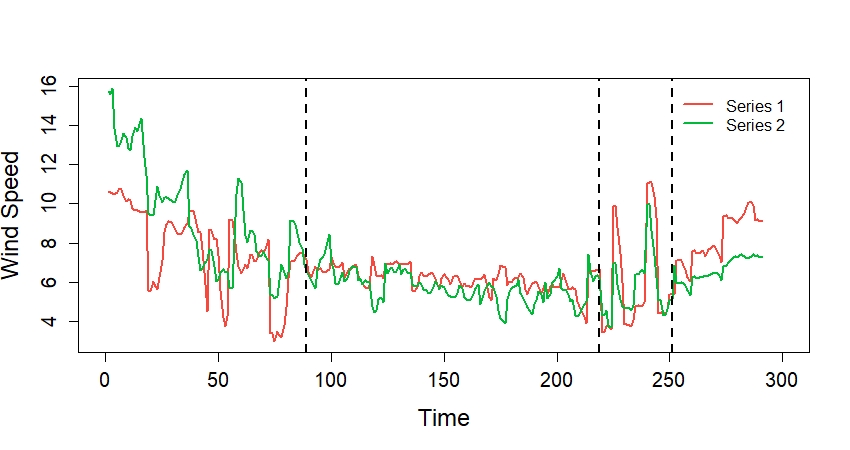}
    \caption{Simulated wind speed.}
    \label{fig:simulatedSeriesSep15}
    \end{subfigure}
    \begin{subfigure}[b]{\columnwidth} 
     \includegraphics[width=\textwidth, height=5cm]{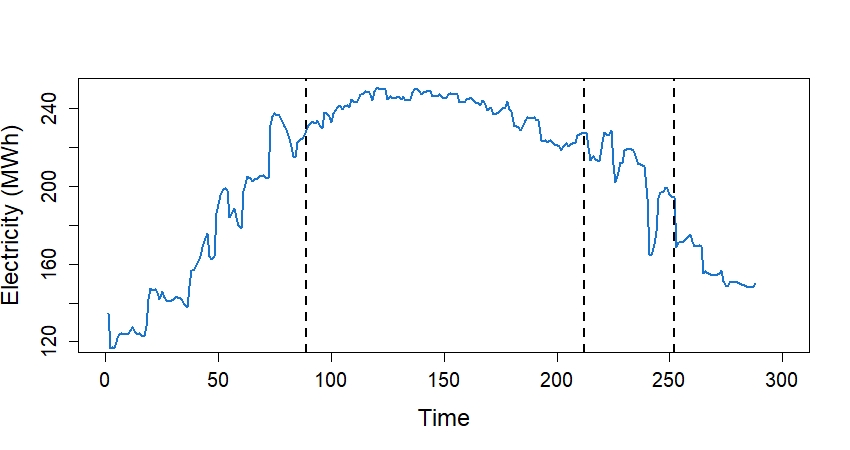}
    \caption{Total conventional generation.} 
    \label{fig:totConGenSep15}
    \end{subfigure}
    \begin{subfigure}[b]{\columnwidth}  \includegraphics[width=\textwidth, height=5cm]{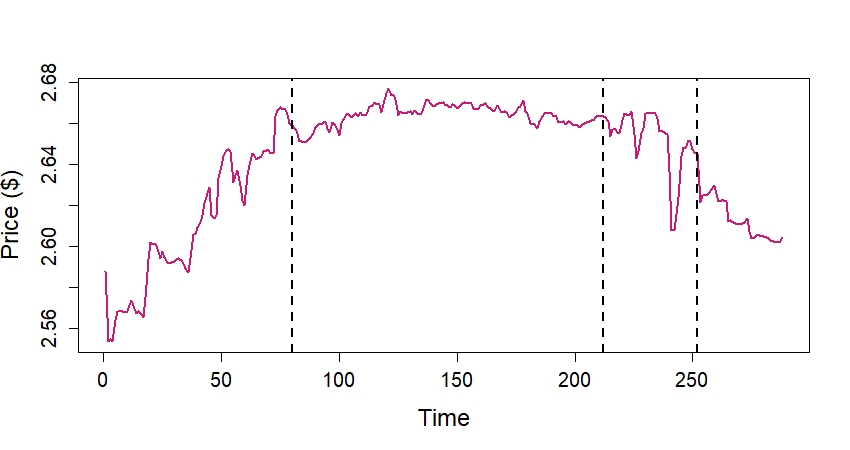}
    \caption{Location Marginal Price at Bus-1}
    \label{fig:priceBus1Sep15}
    \end{subfigure}
    \caption{Results from the Rolling Horizon Economic Dispatch model.}
    \label{fig:clearModelResultSep15}
\end{figure}

The bivariate time series generated using the nonparametric simulation method is shown in Figure \ref{fig:simulatedSeriesSep15}. The correlation between the simulated time series is $0.6664$. This indicates that the simulated time series retain the original linear relationship between the two variates. The change point method detected $t = 89, 219, 251$ as the change points (see Figure \ref{fig:simulatedSeriesSep15}) for the simulated time series indicating that the simulation preserves the structural properties of the original series. Note that the asymptotic properties of change point detection method are well established, however, such behavior can only be observed when a very large number of observations are available. In small sample settings, as is our case, the method searches in a neighborhood of a point for potential locations for change points. Therefore, we expect to see mild discrepancies in change points between the original (e.g., $t = 87$) and the simulated ($t = 89$) time series. 

The wind generation corresponding to the simulated time series was input to the dispatch model \eqref{eq:statModelClearingModel}. Notice that the input time series affects only the upper bound constraint \eqref{eq:statgenBounds} corresponding to the wind generators.  As wind generation is the cheapest out of all the generators in the system, wind generation is fully utilized, when possible, and the conventional generators are adjusted to meet the remaining demand. The total conventional generation is depicted in Figure \ref{fig:totConGenSep15}.  The change point analysis on the total conventional generation time series reveals three change points at $t = 89, 212,$ and $252$. These change points are in the neighborhood of the change points identified for the simulated time series in Figure \ref{fig:simulatedSeriesSep15}. 

In addition to the primal optimal solution (conventional generation, etc.), we also monitored the dual optimal solutions of the optimization model in \eqref{eq:statModelClearingModel}. Since line capacities are considered in the dispatch model, we observed some load-shedding in some buses. Load-shedding is penalized in the objective, and therefore, the prices at these buses are distorted by the load-shedding penalty. We conducted the change point analysis on the location marginal price time series at buses whose marginal prices was not distorted by the load-shedding penalty. For example, the marginal price time series and the detected change points at Bus-1 are depicted in Figure \ref{fig:priceBus1Sep15}.  Change point analysis reveals similar change points (at $ t = 80, 212,$ and $252$) for the price time series as in the input wind speed and the primal total conventional generation times series. 

The primal and dual solutions to quadratic programs with affine constraints, such as \eqref{eq:statModelClearingModel}, are known to be piecewise continuous in the right-hand side parameters. Since wind generation impacts the upper bound constraint in \eqref{eq:statgenBounds}, the observations in Figure \ref{fig:totConGenSep15} and \ref{fig:priceBus1Sep15} provide an empirical evidence for the piecewise relationship of quadratic programming solutions and its input. Furthermore, the neighborhood around the detected change points correspond to intervals that exhibit significant fluctuation in generation and prices. Therefore, the change point detection method can be used for power systems operation and market stability analyses. The piecewise relationship also allows the system operator to predict intervals of volatility, at desired level of significance, by conducting the change point analysis on a wind speed forecast time series.

\section{Conclusions} \label{sec:statConclussion}
In this paper, we emphasized on two principal elements of wind speed modeling, viz., spatio-temporal correlation and nonstationarity. Analyzing wind speed at geographically separated wind farms in isolation (univariate analysis) fails to capture the spatial correlations. We demonstrated the presence of the nonstationarity in both hourly and sub-hourly wind speed time series. Therefore, wind speed time series cannot be directly modeled by usual time series modeling methods such as VAR. Ignoring spatial correlation and nonstationarity restricts the capabilities of power systems planning and operations models to realistically capture the stochastic processes that affect them. 


To address this issue, we presented a change point detection method for multivariate nonstationary wind speed time series. The change point detection method allowed the nonstationary time series to be decomposed into stationary segments. The change point method  sequentially identifies the time points where significant changes in the covariance structure occur. The presented change point detection does not make any distributional assumptions and is supported by sound asymptotic consistency properties.

Once the stationary segments are identified, parametric and nonparametric approaches were presented to simulate new wind speed time series. Generally, the parametric method provides better results if the time series segments satisfy the distributional assumptions of the parametric model. However, parametric models can get computationally expensive and require large amount of data for estimating the model parameters when the model order increases. Therefore, for wind speed data we recommend a nonparametric approach if the parametric model order is greater than five for at least one stationary segment. We presented a block bootstap technique as a reliable nonparametric approach to simulate wind speed time series. 

Finally, we demonstrated the application of change point detection method and subsequent simulation techniques to the economic dispatch problem. We showed that the change points of the wind speed time series correspond to time intervals where the conventional generation and electricity prices also exhibit significant volatility. Such relationship between input time series (wind speed) and optimization model output (generation amounts and electricity prices) can play a critical role in detecting and predicting system instabilities. A thorough study involving stability analysis of power systems, including frequency and voltage stability, from the perspective of change point analysis is a promising research direction. Such an analysis must also take into consideration the renewable generation portfolio comprising of both wind and solar resources. These endeavors will be pursued in our future research. 


\bibliographystyle{unsrt}
\bibliography{statModel_preprint}

\end{document}